\documentclass[11pt]{article}
\usepackage{amsfonts}
\addtocounter{page}{0}
\usepackage[german]{babel}
\begin{document}

\begin{center}
How good is the Warnsdorff's knight's tour heuristic?\\

\vskip .3in

Samuel L. Marateck\\

Department of Computer Science,\\ 
New York University,\\
New York, N.Y.10012,\\
USA\\

\end{center}

\vskip .3in

Warnsdorff's rule for a knight's tour is a heuristic, i.e., it's a rule
that does not produce the desired result all the time. 
It is a classic example of a greedy method in that it is based on a series of locally 
optimal choices. This note describes an
analysis that determines how good the heuristic is on an 8 X 8 chessboard. The
order of appearance in a permutation of the eight possible moves a knight can
make determines the path the knight takes. A computer analysis is done of the
8! permutations of the order of a knight's moves in Warnsdorff's rule on an 8
X 8 chessboard for tours starting on each of the 64 squares. Whenever a tie
occurs for moves to vertices that have the lowest degree, the first of these
vertices encountered in the programming loop is chosen. The number of
permutations of the 8! total that yield non-Hamiltonian paths is tallied. This
will be the same value if we consistently choose the last of these vertices
encountered.\\

\vskip .2in

\begin{enumerate}
\item {\bf Introduction}

The Knight's Tour is an example of a Hamiltonian path on a graph in that the
knight touches each square once and only once on the chessboard. A path in
which the starting square is reachable from the last square the knight visits
is called {\it closed} or {\it re-entrant} or a {\it Hamiltonian cycle}. In
1823 Warnsdorff [5] published a rule to produce a Hamiltonian path: {\it At
each square a look-ahead is performed to see which of the possible next squares
has the least number of valid exits. The square having the least number of
valid exits} -- in graph theory this translate to the vertex with the lowest
degree -- {\it is the next square the knight lands on.  If there is a tie
between two or more squares, the next square is arbitrarily chosen.}  The
look-ahead for the penultimate square (the 63rd one) in a Hamiltonian path
should indicate the last square (the 64th one) has degree zero, i.e., there is
no square to which the knight on the ultimate square can move.

\item {\bf Examples of Knight's tours}

The knight's moves are described by the graph {\it G = (V, E)}, where the
squares on the board are the vertices {\it V} and the allowed moves of the
knight are the edges {\it E}, i.e., $V = \{(i, j) | 1 \le i, j \le 8\}$ and $E
= \{((i,j), (k,l)) | \{|i - k|, |j - l|\} = \{1,2\}\}$.  See for instance [3]
[1] [2] [4].  We describe the change in the knight's x and y positions by the
pair $\langle dx, dy\rangle$. If we use the the following order for the
knight's eight possible moves, $\langle 1,2\rangle \langle 2,1\rangle \langle
1,-2\rangle \langle 2,-1\rangle \langle -1,2\rangle \langle -2,1\rangle \langle
-1,-2\rangle \langle -2,-1\rangle$ to determine which square the knight
procedes to next, if he starts at (0,0)\footnote{We use the convention that the
numbering of the rows and columns of a matrix starts with zero.} the path is
described in Fig. 1.  The path is, incidentally, a closed path.

\newpage
\begin{verbatim}
 1  4 61 20 41  6 43 22
34 19  2  5 60 21 40  7
 3 64 35 62 37 42 23 44
18 33 48 57 46 59  8 39
49 14 63 36 55 38 45 24
32 17 56 47 58 27 54  9
13 50 15 30 11 52 25 28
16 31 12 51 26 29 10 53
\end{verbatim}

\noindent
Fig. 1. The Hamiltonian path starts at (0,0).

The tour in Warnsdorff's algorithm stops when a square with zero exits is
reached. If that square occurs before the knight reaches the 63rd square, as it
does in Fig. 2. at the square at (5,0) the tour halts and the path descibed is
non-Hamiltonian. This tour uses the same $\langle dx, dy\rangle$ permutation as
in Fig. 1. Remember if a few squares are left and one of them having degree
zero is occupied by the knight, the other squares -- they are marked in
Fig. 2. by a zero -- cannot reach the one last occupied. Why? Since a
zero-degree square cannot acess any other squares, no square can access it.\\

\begin{verbatim}
  0  2 19 24 35 28 17 26
 20 23  0  1 18 25 34 29
  3  0 21 36 45 32 27 16
 22 55 46  0 42 37 30 33
 47  4 59 54 31 44 15 38
 60 53 56 43 50 41 12  9
  5 48 51 58  7 10 39 14
 52 57  6 49 40 13  8 11
\end{verbatim}

\noindent
Fig. 2. A non-Hamiltonian path for which the program uses the same $\langle dx,
dy\rangle$ permutation as is used in Fig. 1. but starts at square (1,3).

\item {\bf Analysis of the 8! permutations of the knight's moves}

For each of the 8! or 40,320 permutations of the knight's moves, a program
determines if a tour starting from any of the 64 squares produces a
non-Hamiltonian path. The result is that 32,944 permutations out of 40,320
produce at least one non-Hamiltonian path. So you only have an 18\% chance of
randomly choosing a permutation that will yield a board without any
non-Hamiltonian paths. The most non-Hamiltonian paths resulting from a
permutation is 9. The permutation
\begin{verbatim}
<1,2> <1,-2> <-2,-1> <2,-1> <-2,1> <-1,-2><-1,2> <2,1>
\end{verbatim}
is one of those that produces 9 non-Hamiltonian paths. On the other hand,
\begin{verbatim}
<1,2> <2,1> <1,-2> <-1,2> <-2,-1> <2,-1> <-1,-2> <-2,1>
\end{verbatim}
is one of those that produces 64 Hamiltonian paths and thus no non-Hamiltonian
paths. 

In case of a tie occurring for moves to the two or more vertices having the
same lowest degree, the program chooses the first vertex encountered. The same
result for the sum of the non-Hamiltonian paths would occur if you choose the
last vertex encountered. The reason is that summing over all the possible
permutations makes the two non-Hamiltonian sums -- one obtained using the first
vertex encountered in a tie, and the other using the last vertex encountered --
identical because of the symmetry of the permutations: For every permutation
of the form abcdefgh there is a corresponding one of the form hgfedcba. In an
analysis of the tours themselves, of the 64 x 40,320 or 2,580,480 possible
tours, 78,832 are non-Hamiltonian. Again, the results are independent of
whether the first or last vertex is chosen in a tie.

\end{enumerate}

\noindent
{\bf References}\\
\\
\footnotesize{
\noindent
[1] Axel Conrad, Tanja Hindrichs, Hussein
Morsy, Ingo Wegener, Solution of the Knight's Hamiltonian path problem on
chessboards, Discrete Applied Mathematics 50 (1994) 125-134.

\noindent
[2] Olaf Kyek, Ian Parberry, Ingo Wegener, Bounds on the number of knight's
tours, Discrete Applied Mathematics 74 (1997) 171-181.

\noindent
[3] Ian Parberry, An efficient algorithm for the Knight's tour problem,
Discrete Applied Mathematics 73 (1997) 250-260.

\noindent
[4] Ira Pohl, A Method for Finding Hamiltonian Paths and Knight's Tours,
Communications of the ACM 10 (1967) 446-449.

\noindent
[5] H. C. Warnsdorff, Des R$\ddot{o}$sselsprunges einfachste und
allgemeinste L$\ddot{o}$sung. Schmalkalden, (1823).
}

\end{document}